\documentclass[fleqn,usenatbib]{mnras}
\usepackage[T1]{fontenc}
\usepackage{ae,aecompl}
\usepackage{graphicx}	
\usepackage{booktabs}
\usepackage{microtype}
\usepackage{multirow}
\usepackage{tikz}
\usetikzlibrary{calc}
\usepackage{mathptmx}
\usepackage{amsmath}	
\usepackage{amssymb}	
\usepackage[modulo, switch]{lineno}
\usepackage{listings}
\newcommand{\new}[1]{{#1}}
\newcommand{\twocell}[1]{\multicolumn{2}{c}{#1}}
\newcommand{\starA}{KIC~8410637}
\newcommand{\starB}{KIC~9540226}
\newcommand{\starC}{KIC~5640750}
\newcommand{\BV}{Brunt--V{\"a}is{\"a}l{\"a}}
\newcommand{\st}[1]{_\mathrm{#1}}
\newcommand{\FeH}{[\mathrm{Fe}/\mathrm{H}]}
\newcommand{\Teff}{T\st{eff}}
\newcommand{\K}{\,\mathrm{K}}
\newcommand{\yr}{\,\mathrm{yr}}

\newcommand{\uHz}{\,\mu\mathrm{Hz}}
\newcommand{\tdif}[2]{\frac{\mathrm{d}#1}{\mathrm{d}#2}}
\newcommand{\eq}[1]{
\begin{equation}
#1
\end{equation}}

\title{Surface effects on the red giant branch}
\author[W. H. Ball et al.]{
W. H. Ball,$^{1,2}$\thanks{E-mail: wball@bison.ph.bham.ac.uk}
N. Theme{\ss}l$^{3,2,4}$
and S. Hekker$^{3,2}$
\\
$^{1}$School of Physics and Astronomy, University of Birmingham, Edgbaston, Birmingham B15 2TT, United Kingdom\\
$^{2}$Stellar Astrophysics Centre, Department of Physics and Astronomy, Aarhus University, Ny Munkegade 120, DK-8000 Aarhus C, Denmark\\
$^{3}$Max-Planck-Institut f{\"u}r Sonnensystemforschung, Justus-von-Liebig-Weg 3, D-37077 G{\"o}ttingen, Germany\\
$^{4}$Institut f{\"u}r Astrophysik, Georg-August-Universit{\"a}t G{\"o}ttingen, Friedrich-Hund-Platz 1, 37077 G{\"o}ttingen, Germany
}

\date{Accepted XXX. Received YYY; in original form ZZZ}

\pubyear{2018}
\pagerange{\pageref{firstpage}--\pageref{lastpage}}

\begin{document}

\label{firstpage}
\maketitle

\begin{abstract}
Individual mode frequencies have been detected in thousands of
individual solar-like oscillators on the red giant branch (RGB).
Fitting stellar models to these mode frequencies, however, is more
difficult than in main-sequence stars.  This is partly because of the
uncertain magnitude of the surface effect: the systematic difference
between observed and modelled frequencies caused by poor modelling of
the near-surface layers.  We aim to study the magnitude of the surface
effect in RGB stars.  Surface effect corrections used for
main-sequence targets are potentially large enough to put the
non-radial mixed modes in RGB stars out of order, which is unphysical.
Unless this can be circumvented, model-fitting of evolved RGB stars is
restricted to the radial modes, which reduces the number of available
modes.  Here, we present a method to suppress gravity modes (g-modes)
in the cores of our stellar models, so that they have only pure
pressure modes (p-modes). We show that the method gives unbiased
results and apply it to three RGB solar-like oscillators in
double-lined eclipsing binaries: \starA{}, \starB{} and \starC{}.  In
all three stars, the surface effect decreases the model frequencies
consistently by about $0.1$--$0.3\uHz$ at the frequency of maximum
oscillation power $\nu\st{max}$, which agrees with existing
predictions from three-dimensional radiation hydrodynamics
simulations.  Though our method in essence discards information about
the stellar cores, it provides a useful step forward in understanding
the surface effect in RGB stars.
\end{abstract}

\begin{keywords}
  binaries: eclipsing -- stars: interiors -- stars:
  evolution -- stars: oscillations -- stars: individual: \starA{}, \starB{}, \starC{}
\end{keywords}

\section{Introduction}

Long-term space-based monitoring---principally from CoRoT
\citep{corot} and \emph{Kepler} \citep{kepler}---has led to the
detection of oscillation mode frequencies in thousands of
solar-like oscillators.  Most of these
solar-like oscillators are in
the helium core-burning red clump (RC) or on the
hydrogen shell-burning red giant branch (RGB), where their
oscillations are slower (and
therefore detectable at longer cadences), and many major
advances and discoveries have been made using these data \citep[see
  e.g.][for a review]{hekker2017}.  An early breakthrough was in the
use of the period spacing between dipole mixed modes to distinguish RC
stars from RGB stars, even when their non-seismic observables (e.g.
surface gravity $\log g$ and effective temperature $\Teff$) make them
hard to tell apart \citep{beck2011, bedding2011}.  This analysis is
now common, and has been used to identify smaller subgroups, including
stars in the secondary clump or potentially going through the helium
flash \citep{mosser2014}.  A second major discovery has been the
measurement of core rotation rates in subgiants and red giants
\citep{beck2012, deheuvels2012}.  These core rotation rates are slower
than current models predict \citep[e.g.][]{marques2013} and
have sparked research into other mechanisms that could transport
angular momentum between the core and envelope
\citep[e.g.][]{belkacem2015a,belkacem2015b}.  Finally, given that they
can be seen at distances of thousands of parsecs, red giants have
contributed to multiple results in the field of galactic archaeology
\citep[e.g.][]{miglio2013}.

Despite the progress outlined above, there has been less success in
fitting stellar models directly to the observed mode frequencies, as
is now relatively routine for main-sequence solar-like oscillators
\citep[e.g.][]{appourchaux2015, white2017, legacy2, creevey2017}.
This is partly because red giants take more
time to model: the evolutionary tracks must evolve further, the
evolutionary computations become slower, and the dense spectrum of
non-radial modes takes much longer to compute, though some progress is
being made.  \citet{perez2016} fit stellar models to the frequencies
measured by \citet{corsaro2015} by using the period spacing
between dipole ($\ell=1$) mixed modes and the individual mode
frequencies for the radial, quadrupole and octupole modes
($\ell=0,2,3$).  More recently, \citet{li2018} fit stellar models to a
sample of red giants in eclipsing binaries using the stellar models to
guide the identification of mode frequencies in the observations.

In addition to the slower computation, there is the problem of the `surface effect': the
systematic difference between observed and modelled frequencies caused
by poor modelling of the near-surface layers \citep[see][for a recent review]{ball2017kasc}.
The effect is
well-known in the Sun whereas it is hard to know what to
expect in red giants.  \citet{sonoi2015} computed the frequency shifts
caused by replacing the near-surface layers of stellar
models---including a red giant---but restricted the calculation to
radial modes.

For non-radial modes, the frequency shifts are potentially larger than
the observed period spacings between the many mixed modes, in which
case the parametrisations used for dwarfs would imply an unphysical
breaking of the ordering of the modes.
If all neighbouring modes were shifted by similar amounts (as is
  in dwarf solar-like oscillators), this would not be a problem.
  For mixed modes, however, we expect the surface effect to be greater in those
  with a stronger p-mode component.  This can mean that two neighbouring modes are shifted by
  very different amounts (see Sec.~\ref{ss:nog} and
  Fig.~\ref{f:badcorr}).  What we really expect is that the
  p-mode component is shifted by the surface effect and the shifted p-mode
  couples to a different
  g-mode in the core.  Our study of the surface effect would still be
  possible if we restrict our attention to radial modes only, for
  which mixed modes are not possible.  This, however, would come at
  the cost of discarding about two thirds of our observed p-dominated mode
  frequencies and reduce the precision of our parameters.  A simplistic
  calculation using the results presented here suggests that our
  derived uncertainties would be roughly 70 per cent larger if
  we only used the radial mode frequencies.

To exploit the non-radial mode frequencies, we isolate the p-mode
components of the mixed modes by suppressing the g-mode oscillations.
We achieve this by setting the squared \BV{} frequency $N^2$ to zero
throughout the convectively-stable core (see Sec.~\ref{ss:nog} for
details).  We note that this is one of several ways to extract
``pure'' p-mode frequencies.  For example, one alternative is to
restrict the oscillation calculation for the non-radial modes to the
convective envelope.  We have chosen to set $N^2$ to zero because it
is straightforward to implement.\footnote{Our implementation in MESA
  requires \new{13} lines of additional code \new{(see
    Appendix~\ref{a:nog})}.}

The complementary information available for stars in binary
systems makes them ideal targets for constraining stellar physics.
In particular, double-lined eclipsing
binaries (DEBs) allow independent measurements of the masses and radii
of the components.  There are few main-sequence solar-like
  oscillators in binaries, as predicted by \citet{miglio2014}.
Given their greater number, we expect more binaries
containing RGB or RC stars, though usually only one component will
have measurable oscillations.  Nearly 20 such systems are now known
and they have been used to study potential biases in the asteroseismic
scaling relations \citep[e.g.][and references
    therein]{gaulme2016,themessl2018}.

Here, we fit stellar models to mode frequencies measured for the RGB
stars in three DEBs, with the aim of investigating whether
the surface corrections used for main-sequence solar-oscillators are
still valid and agree with our expectations.  As a secondary result,
we also compare the stellar parameters that are recovered with those
found from dynamical modelling of the binary system.

\begin{table*}
  \centering
  \caption{Table of stars' observed properties, taken from
      \citet{themessl2018}.}  \begin{tabular}{ccccccccccccccccccccccccccccccc}
\toprule
Star & KIC & $\Teff/\K$ & $\FeH$ & $\nu\st{max}/\uHz$ & $\Delta\nu/\uHz$ &
$M/M_\odot$ & $R/R_\odot$ \\ 
\midrule
A & 8410637 & $4605 \pm 80$ & $\phantom{-}0.02 \pm 0.08$ & $46.4 \pm 0.3$ & $4.564 \pm 0.004$ &
$1.472 \pm 0.017$ & $10.596 \pm 0.049$\\
B & 9540226 & $4585 \pm 75$ & $-0.31 \pm 0.09$ & $26.7 \pm 0.2$ & $3.153 \pm 0.006$ &
$1.390 \pm 0.031$ & $13.43 \pm 0.17$ \\
C & 5640750 & $4525 \pm 75$ & $-0.29 \pm 0.09$ & $24.1 \pm 0.2$ & $2.969 \pm 0.006$ &
$1.158 \pm 0.014$ & $13.119 \pm 0.090$ \\
\bottomrule
\end{tabular}

  \label{t:obs}
\end{table*}

\begin{table}
  \centering
  \caption{Table of stars' orbital properties, taken from
      \citet{themessl2018}.}  \begin{tabular}{ccccccccccccccccccccccccccccccc}
\toprule
Star & $P/\mathrm{d}$ & $e$ & $q$ \\
\midrule
A & $408.3248 \pm 0.0004$ & $0.694 \pm 0.004$ & $0.890 \pm 0.005$ \\
B & $175.4438 \pm 0.0008$ & $0.387 \pm 0.003$ & $0.730 \pm 0.032$ \\
C & $987.398  \pm 0.006$  & $0.322 \pm 0.008$ & $0.971 \pm 0.012$ \\
\bottomrule
\end{tabular}

  \label{t:orb}
\end{table}

\begin{figure*}
  \centering
\begin{tikzpicture}
  \node[inner sep=0pt] (img) at (0,0)
       {\includegraphics[width=2\columnwidth]{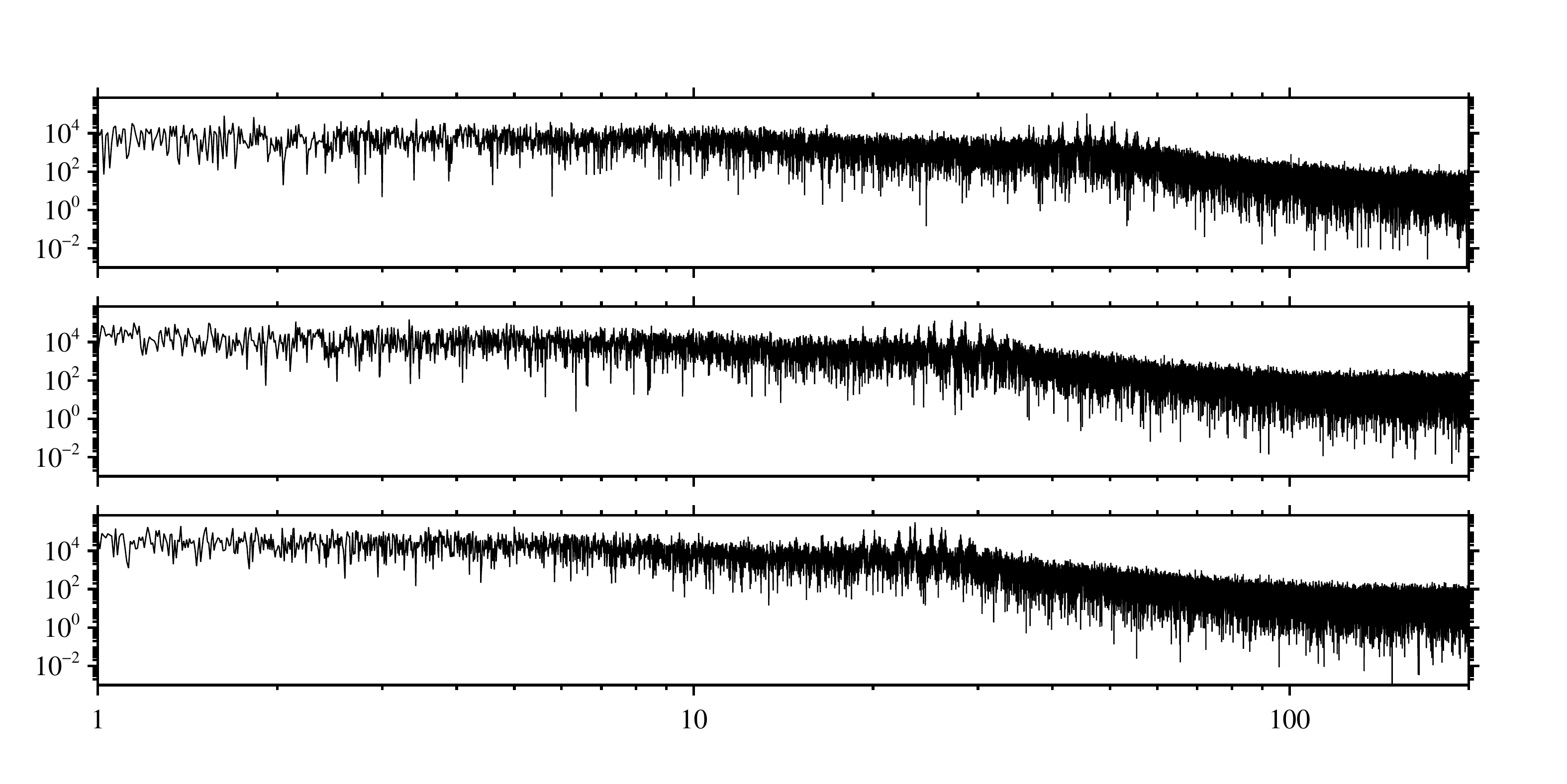}};
  \node (x) at ($(img.south)!0.02!(img.north)$) {\Large $\nu/\uHz$};
  \node[rotate=90] (y) at ($(img.west)!0.0!(img.east)$) {\Large PSD$/(\mathrm{ppm}^2/\uHz)$};
  \node (A) at (-6.1,  1.85) {\large \starA{}};
  \node (B) at (-6.1, -0.4) {\large \starB{}};
  \node (C) at (-6.1, -2.65) {\large \starC{}};
 \end{tikzpicture}
 \caption{Power density spectra of \starA{} (star A, top), \starB{} (star B, middle)
   and \starC{} (star C, bottom).}
 \label{f:power_spectra}
\end{figure*}

\section{Observations, models and fitting method}
\label{s:models}

\subsection{Target stars}

Our asteroseismic targets are the three red giants \starA{}, \starB{}
and \starC{}, hereafter referred to as stars A, B and C, observed by
\emph{Kepler} during its nominal mission.  Power spectra for the
three stars are shown in Fig.~\ref{f:power_spectra} and their basic stellar parameters
and orbital parameters
\citep[from][]{themessl2018} are
listed in Tables~\ref{t:obs} and \ref{t:orb}.  All three stars have been identified
as RGB stars and are parts of
detached, eclipsing, spectroscopic binary systems that have been
studied since their respective discoveries.  Star A was the first
oscillating red giant detected in an eclipsing binary
\citep{hekker2010}.  \citet{frandsen2013} subsequently obtained
high resolution spectroscopy from which they measured the masses and
radii of both components.  \citet{beck2014} analysed a sample of 18 red
giants in eccentric binary systems, including star B.
\citet{brogaard2016} also analysed stars A and B.
All three stars were
part of the ensembles studied by \citet{gaulme2013, gaulme2014}, and
\citet{gaulme2016} included stars A and B in their comparison of
dynamical and asteroseismic masses.

The data used here are from \citet{themessl2018}, who measured the
oscillation frequencies of stars A, B and C, as well as computing
updated orbital solutions to the light curves and radial
velocities.  They compared the masses and radii from the orbital
solutions with those produced by asteroseismic scaling relations
\citep{brown1991, kjeldsen1995} and grid-based modelling, and found that the
results agree if the scaling relations are corrected for variations
with mass, temperature, metallicity and surface effects.
\citet{themessl2018} report two masses
  and radii for star C derived from the eclipse and radial velocity
  observations.  The two orbital solutions are of similar quality,
  with the ambiguity principally caused by poor coverage of the
  radial velocity observations as a function of orbital phase.
  We initially considered both solutions and,
  like \citet{themessl2018}, concluded that the lower-mass solution
  was consistent with our seismic results for the other two stars.
  To avoid confusion, we hereafter restrict our attention to the
  lower-mass orbital solution for star C.

\subsection{Stellar models}

We computed stellar models using the Modules for Experiments in
Stellar Astrophysics
\citep[MESA\footnote{\url{http://mesa.sourceforge.net}}, revision
  9575,][]{paxton2011,paxton2013,paxton2015}. Opacities at high and low
temperatures are taken from the tables by the OPAL collaboration
\citep{iglesias1996} and \citet{ferguson2005}, respectively.  Nuclear
reaction rates are drawn either from the NACRE tables
\citep{angulo1999} or, if a given rate was not available there, from
the tables by \citet{caughlan1988}.  For the specific reactions
${}^{14}\mathrm{N}(p,\gamma){}^{15}\mathrm{O}$ and
${}^{12}\mathrm{C}(\alpha,\gamma){}^{16}\mathrm{O}$, we use the
revised rates by \citet{imbriani2005} and \citet{kunz2002}.
Convection is described by mixing-length theory
\citep{boehm-vitense1958} as derived in \citet{cox1968}
with no overshooting.  For the
solar abundances, we use the overall abundance and mixture given by
\citet{grevesse1998}.  For the surface boundary condition, we extended
the outermost meshpoint of the stellar model to an optical depth of
$\tau=10^{-4}$, which in effect creates a grey Eddington atmosphere
that is included in the interior model.  Photospheric values are
determined by interpolating at the photospheric optical depth
$\tau=2/3$.

We note that our models omit the effects of
gravitational settling, radiative levitation and rotation.
The three stars here are massive enough that the current
  implementation of gravitational settling would completely drain the
  surface of helium and metals during the main sequence evolution, which we
  regard as less justified than omitting gravitational settling.
  In MESA, calculations including radiative levitation are presently too
  time-consuming and rotation is only implemented in the
  diffusion approximation, so we omit them here.
Mode frequencies were computed using the Aarhus adiabatic pulsation
code \citep[ADIPLS,][]{adipls} without remeshing.

\begin{figure}
  \centering
\begin{tikzpicture}
  \node[inner sep=0pt] (img) at (0,0) {\includegraphics[width=\columnwidth]{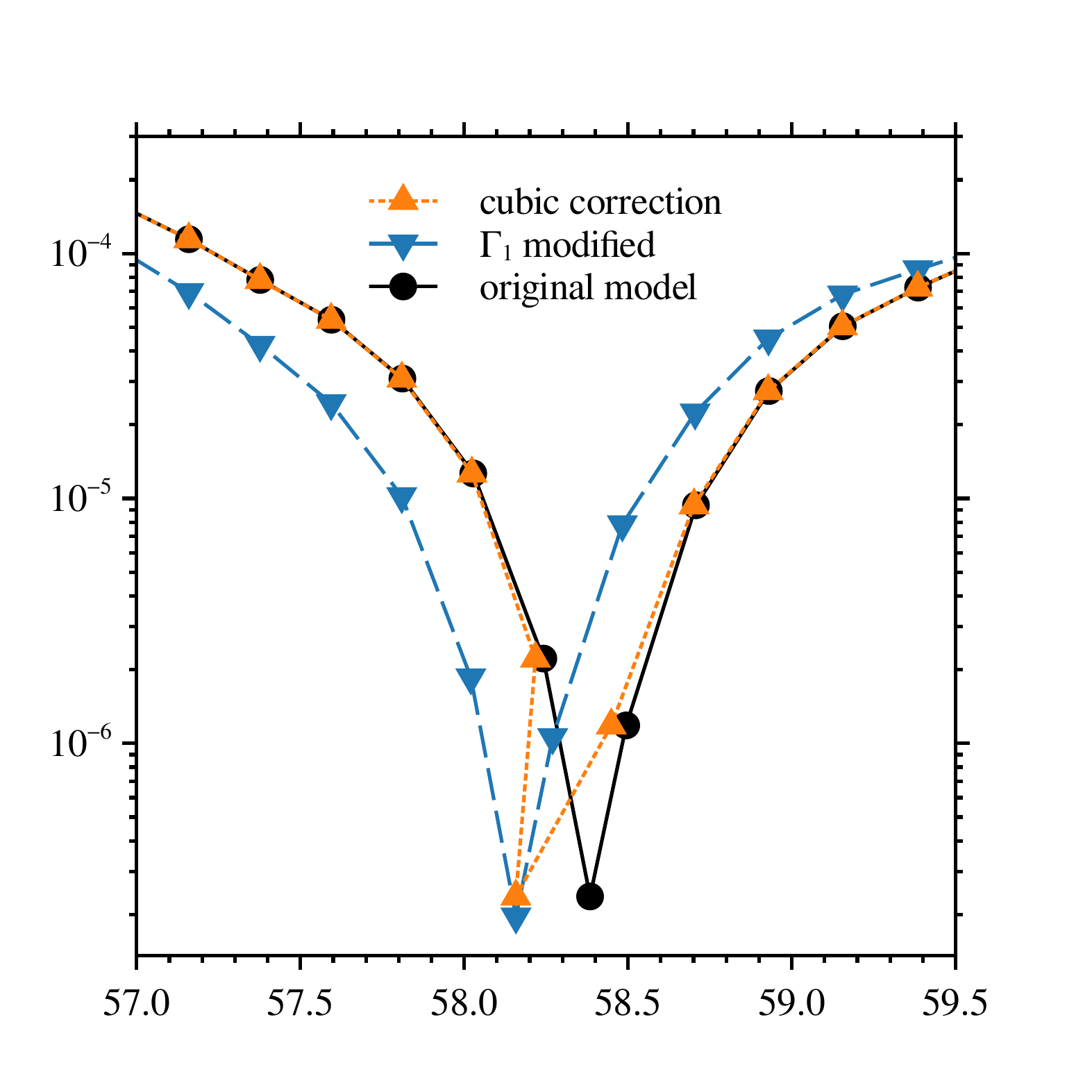}};
  \node (x) at ($(img.south)!0.02!(img.north)$) {$\nu/\mu\mathrm{Hz}$};
  \node[rotate=90] (y) at ($(img.west)!0.02!(img.east)$) {$\mathcal{I}$};
 \end{tikzpicture}
\caption{Mode inertiae as a function of frequency for the stellar
  model used as a hare for \starA{} (see the end of this section).
  The solid black curve shows the values
  for the unmodified model, the dashed blue curve for a model in which the
  first adiabatic index $\Gamma_1$ has been decreased near the
  surface, and the dotted orange curve the original values with a cubic
  surface correction applied to try to match the mode with the lowest
  inertia in the modified model (blue).
  In the corrected frequencies (orange), the correction is
  large enough to change the order of the modes with frequency.}
 \label{f:badcorr}
\end{figure}

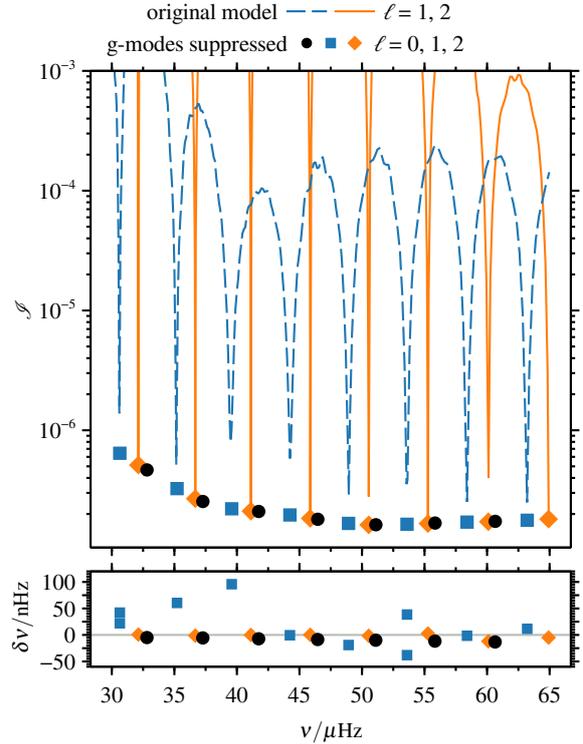
\begin{figure}
  \centering
\begin{tikzpicture}
  \node[inner sep=0pt] (img) at (0,0) {\includegraphics[width=\columnwidth]{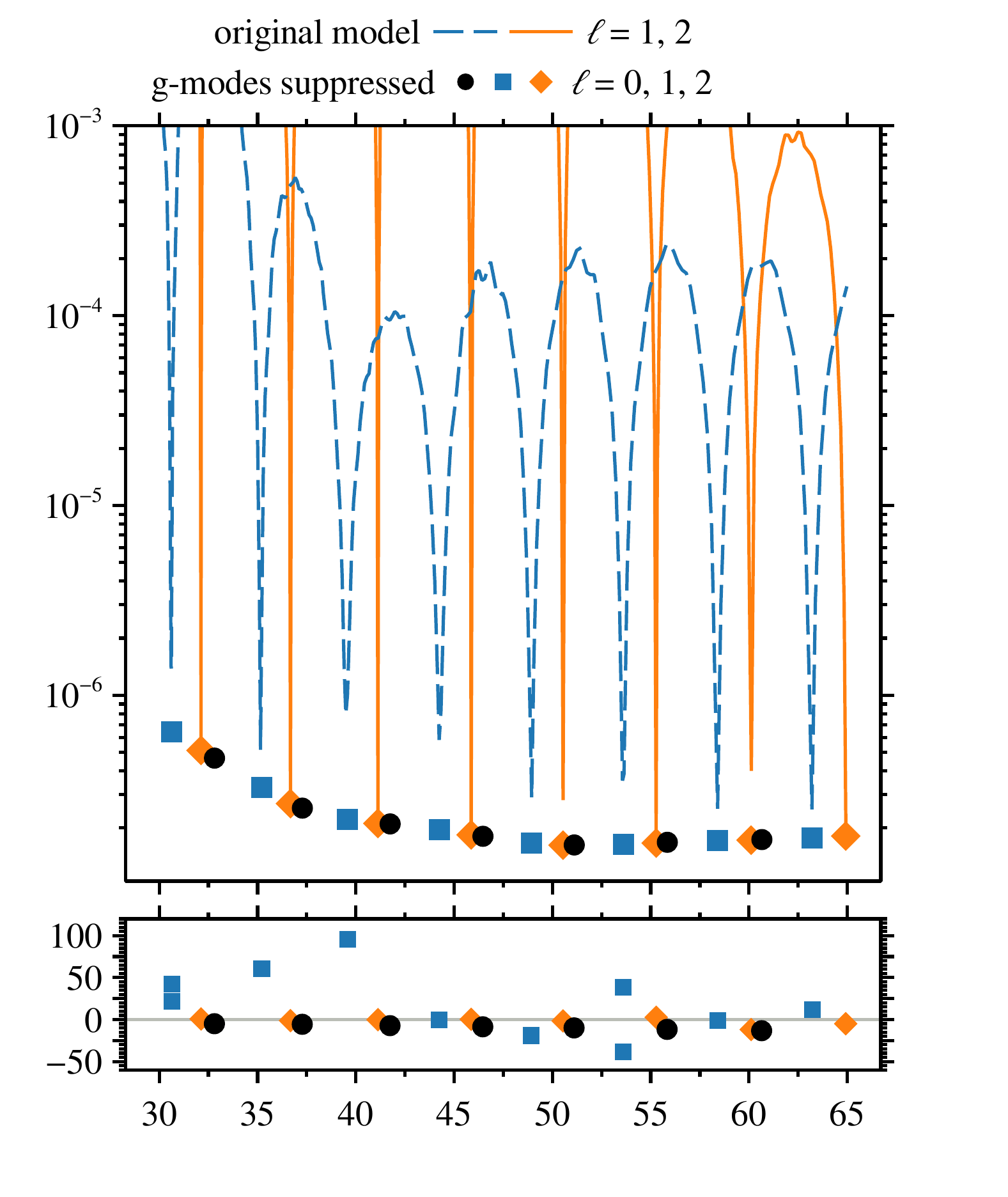}};
  \node (x) at ($(img.south)!0.02!(img.north)$) {$\nu/\mu\mathrm{Hz}$};
  \node[rotate=90] (y1) at ($(img.west)!0.16!(img.north west)!0.02!(img.east)$) {$\mathcal{I}$};
  \node[rotate=90] (y2) at ($(img.west)!0.67!(img.south west)!0.02!(img.east)$) {$\delta\nu/\mathrm{nHz}$};
 \end{tikzpicture}
\caption{Demonstration of the effect of suppressing g-mode
  oscillations, as described in Sec.~\ref{ss:nog}, in the model of
  star A (\starA{}) that served as the hare.  In the upper panel,
  the dashed blue and
  solid orange lines show the mode inertiae as a function of frequency
  for the $\ell=1$ and $2$ modes before the g-modes were
  suppressed. The black circles, blue squares and orange diamonds show
  the $\ell=0$, $1$ and $2$ modes after the g-modes are suppressed.
  The lower panel shows the differences between the mode
  frequencies before and after the g-modes are suppressed.
  \new{The symbols are as in the upper panel.  For the non-radial
    modes the differences are shown relative to modes with the lowest inertia
    for each radial order in the unmodified model.  For two dipole modes
    (at $30.6$ and $53.6\uHz$), there is another mode with an inertia within
    10 per cent of the lowest, which we have also shown.}}
 \label{f:nog}
\end{figure}

\subsection{Suppressing mixed modes}
\label{ss:nog}

As a solar-like oscillator evolves up the red giant branch, so its
spectrum of mixed modes becomes more closely spaced.  Eventually, the
surface corrections are potentially larger than the observed spacing
between g-modes, in which case the corrections that are used for
dwarfs and subgiants will break down.
This phenomenon is
demonstrated in Fig.~\ref{f:badcorr}, which shows mode inertiae $\mathcal{I}$ as a
function of frequency for the model used as a hare for \starA{} (see
Sec.~\ref{ss:nog}).\footnote{The figure follows the example discussed
  by \citet[][Sec.~4.2]{li2018} and shown in their Figs 3 and 4.}  The
black curve shows the inertiae of the unmodified stellar model.  The
blue curve shows the inertiae in the same model after the
mode frequencies have been shifted by introducing a sound speed
  perturbation concentrated at the stellar surface.  Specifically,
  we modified the first adiabatic index $\Gamma_1$ by assigning it a new value
\eq{\Gamma_1'=\Gamma_1\cdot\left(1-0.4\,e^{-10^6\left[(r/R)-1\right]^2}\right)\label{e:newG1}}
though any sharp peak at the stellar surface will produce a
  similar effect.  Here, $r$ is the radial co-ordinate in the stellar
model and $R$ is the photospheric radius.  The frequencies and
  mode inertiae are changed such that a different mode is now
  identified as the most p-dominated (i.e.~the mode with the lowest
  inertia, see blue curve).  The orange curve shows the original mode frequencies but
now corrected using the cubic correction by \citet{ball2014}
so that the frequency of the most p-dominated mode matches
that of the modified model.  The size of the surface correction
applied to match the black mode to the blue is so
large that it disrupts the monotonic relationship between radial order
and mode frequency (see orange curve).

To address this problem, we
note that the observed frequencies of the three red giants
  in binary systems contain one mode per acoustic radial order and
per angular degree \citep{themessl2018}.  This is always true
for radial modes and in these observations it is also true for
the non-radial modes, even though many mixed modes are theoretically
present.  The observed non-radial modes are those with the
largest amplitudes and presumably the lowest inertiae, which implies
that they are also the least mixed.  If there were no mode mixing,
there would be only one ``pure'' p-mode per acoustic radial
order and per angular degree.  To proceed, we therefore assume that
the observed modes are those whose frequencies are closest to these
hypothetical pure p-modes, and we modify the stellar models to compute
the frequencies of these pure p-modes.  \new{We describe these modifications
  here and provide the Fortran source code in Appendix~\ref{a:nog}.}

We suppress g-modes in the core of the star by setting the square of
the \BV{} frequency $N$ to zero in the
convectively-stable regions beneath the convective envelope, i.e. we
assign $N^2=0$ wherever $N^2>0$ in the core.  In practice, we use the
dimensionless square of the \BV{} frequency
$\mathcal{A}$\footnote{This is sometimes referred to as the
  \emph{Ledoux discriminant} or \emph{convective parameter}.}, which
is defined by
\eq{\mathcal{A}=\frac{r}{g}N^2 =\frac{1}{\Gamma_1}\tdif{\ln P}{\ln
    r}-\tdif{\ln\rho}{\ln r}}
where $r$ is the radial co-ordinate, $g$ the gravity,
$\Gamma_1$ the first adiabatic index, $P$ the pressure and
$\rho$ the density in the unmodified stellar model.

We modify the stellar model by setting the dimensionless
  square of the \BV{} frequency to a new value
  $\mathcal{A}'=0$ wherever $\mathcal{A}>0$ in the core.  We do not
  wish to change the pressure or density, so to keep the model
  consistent with the definition above we must change the first
    adiabatic index to
\eq{\Gamma'_1=\frac{\mathrm{d}\ln P/\mathrm{d}\ln r}
  {\mathrm{d}\ln\rho/\mathrm{d}\ln r}}
Thus, during the model-fitting, we change $\mathcal{A}$ to
  $\mathcal{A}'$ and $\Gamma_1$ to $\Gamma'_1$ in the stellar models
before they are loaded by the oscillation code.

As a practical point, rather than recompute the radial derivatives
above from finite differences, we use existing information in the
stellar model.  The pressure gradient is known from the equation of
hydrostatic equilibrium,
\eq{\tdif{P}{r}=-\rho g}
and the density gradient is provided by the \BV{}
frequency before the stellar model is modified. i.e.
\eq{\tdif{\rho}{r}=\frac{1}{\Gamma_1}\tdif{\ln P}{\ln r}-\mathcal{A}}
where we emphasize that all the variables are from the unmodified
stellar model.
Fig.~\ref{f:nog} shows the resulting change in the mode frequencies
and inertiae in a model of star A.  The upper panel shows that the mode frequencies
  after modifying the model occur at the minima of the mode inertiae,
  where we expect the pure p-modes would occur.  The lower panel
    shows the differences in the mode frequencies before and after
    modification.  \new{The frequency differences for the
      radial modes} should theoretically be zero.  We have modified
    $\Gamma_1$, which affects the radial modes, but the frequencies
    change by less than $0.015\uHz$, which is negligible.

To validate the mode suppression method described above, we performed
a simple hare and hounds exercise using a model of star A produced by
an early iteration of our model-fitting routine.  We computed the full
frequency spectrum of the stellar model (the hare) and then selected
the same modes as the observations.  For the non-radial modes, we used
the modes with the lowest mode inertiae, which gave one mode per
angular degree and radial order.  We perturbed these mode frequencies
(and the non-seismic data) by random variates drawn from normal
distributions with the observed variances and then found a
best-fitting model (the hound) using the mode suppression described
above.

The first two rows of Tables~\ref{t:mdl} and \ref{t:inf} give the
parameters of the target model (the hare) and the best-fitting model
(the hound).  The hound clearly recovers the properties of the hare
within the uncertainties, which demonstrates that the mode suppression
method does not bias our results.

\subsection{Model-fitting procedure}
\label{ss:fit}

We fit the stellar models to the observed data using essentially the
same method as in \citet{ball2017}.  We minimised the total $\chi^2$,
defined by
\eq{\chi^2=\sum_{i=1}^{N\st{obs}}\left(\frac{y_{\mathrm{obs},i}-y_{\mathrm{mdl},i}}{\sigma_i}\right)^2}
where $y_{\mathrm{obs},i}$, $y_{\mathrm{mdl},i}$ and $\sigma_i$ are
the observed value, modelled value and observed uncertainty of the
$i$-th observable, of which there are $N\st{obs}$ in total.  The
observables are the effective temperature $\Teff$, surface metallicity
$\FeH\st{s}$ and the individual mode frequencies.  For each star, we
also recomputed $\chi^2$ including the dynamical mass $M$ and radius
$R$ as constraints.

To correct for the surface effect, we used either the one-term
  (cubic) or two-term (combined) corrections by \citet{ball2014}.
  These fit the differences between the modelled and observed
  frequencies $\delta\nu_i=\nu_{\mathrm{obs},i}-\nu_{\mathrm{mdl},i}$ with the formulae
  \eq{\delta\nu_i = a_3\left(\nu_{\mathrm{mdl},i}/\nu\st{ac}\right)^3/\mathcal{I}_i\label{e:cube}}
  or
  \eq{\delta\nu_i = \left(a_{-1}\left(\nu_{\mathrm{mdl},i}/\nu\st{ac}\right)^{-1}
    +a_3\left(\nu_{\mathrm{mdl},i}/\nu\st{ac}\right)^3\right)/\mathcal{I}_i\label{e:both}}
  for the one- or two-term correction, respectively.  Here, $\nu\st{ac}$
  is the acoustic cut-off frequency, used to non-dimensionalise the equations.
  For convenience, it is computed using the scaling relation
  \eq{\frac{\nu\st{ac}}{\nu_{\mathrm{ac},\odot}}=\frac{g}{g_\odot}
    \left(\frac{T\st{eff}}{T_{\mathrm{eff},\odot}}\right)^{-1/2}}
  and we use $\nu_{\mathrm{ac},\odot}=5000\uHz$, $\log g_\odot=4.438$
  and $T_{\mathrm{eff},\odot}=5777\K$.  The best-fitting values of the
  coefficients $a_{-1}$ and $a_3$ are found by linear regression
  using all the observed modes.
We used the one-term (cubic) correction (eq.~\ref{e:cube}) unless otherwise noted.

We created a handful of initial guesses for each star using the
scaling relations and then proceeded with an iterative method.  For
each choice of mass $M$, initial helium abundance $Y_0$, initial
metallicity $\FeH_0$ and mixing-length parameter $\alpha$, we started
an evolutionary track from a chemically-homogeneous pre-main-sequence
model with central temperature $9\times10^5\K$.  The timestep was
gradually reduced to a minimum of $10^{4.5}\yr$ as the stellar model
first matched the spectroscopic parameters and then the radial mode
frequencies.  Once these requirements were met, the full set of mode
frequencies was computed (using the mode suppression method) and the
total $\chi^2$ recorded.  The parameters for each model with
$\chi^2<2000$ were stored.  As described in more detail below, we
  found best-fitting models both before and after the RGB bump: the
  brief decrease in luminosity as the hydrogen-burning shell passes
  through the composition discontinuity left by the convective
  envelope at its maximum depth.  We used the full sequence of models
  with $\chi^2<2000$ to identify the best-fitting models for each case
  separately.

We then iterated on the parameters of the best-fitting
models---separately for the pre- and post-bump models---principally
using the Nelder--Mead downhill simplex method \citep{nelder1965}.
When the downhill simplex fails, it usually resorts to shrinking the
entire simplex towards the current best-fitting model.  To
avoid this, we tried to produce better-fitting models using a
variety of methods, including: linear extrapolations from subsets of
the sample so far; uniformly-distributed models with the $1$ to
$3\sigma$ confidence regions; extensions of lines between selected
pairs of models; and reflections across the best fitting model.  This
process, though somewhat haphazard, is aimed at preventing convergence
on a local minimum.  The search for better models ended when several
dozen attempts failed to reduce the best $\chi^2$ by more than one.

We determined uncertainties of the model parameters (i.e.~the
  parameters in Table~\ref{t:mdl}, which define the stellar models)
from ellipsoids bounding surfaces of constant $\chi^2$, in particular
by finding the ellipsoids that would simultaneously enclose all the
parameters of all models within an ellipsoid corresponding to that
value of $\chi^2$.  That is, if the minimum of $\chi^2$ was
$\chi^2_0$, we required that models with $\chi^2=\chi_0^2+1$ and
$\chi_0^2+4$ are simultaneously contained within the $1\sigma$ and
$2\sigma$ ellipsoids.  Uncertainties for derived quantities
(i.e.~the quantities given in Table~\ref{t:inf}, which are
  derived from the stellar models) were derived by calculating a
linear fit to each derived property with respect to the model
parameters and propagating the uncertainties linearly.

The method above describes our fiducial fits, denoted ``Fid.'' in
Tables \ref{t:mdl}, \ref{t:inf} and \ref{t:sonoi}.  We also performed
a second set of fits using the two-term surface correction proposed by
\citet{ball2014}, denoted ``BG14-2'' (eq.~\ref{e:both}); a third set in which the
mixing-length parameter was fixed to the solar-calibrated
value of $\alpha_\odot=1.66$, denoted ``$\alpha_\odot$''; and a fourth
set in which the dynamical masses and radii were included as
observational constraints, denoted ``MR''.

\begin{table*}
  \centering
  \caption{Model parameters for the various model fits.  The first two
    rows give the hare and hound by which we tested our g-mode
    suppression method for bias (see Sec.~\ref{ss:nog}).  The
    remaining rows are eight fits for each of stars A, B and C, as
    described in Sec.~\ref{ss:fit}.  The first columns specify
      the runs as described in Sec.~\ref{ss:fit}, grouped by star (A,
      B or C).  The remaining columns give the mass, initial
    helium abundance, initial metallicity, mixing-length parameter,
    age and surface term coefficients for each fit.}
  \begin{tabular}{cr@{}lcccccccccccccccccccccccccc}
\toprule
Star & \twocell{Run} & $M/M_\odot$ & $Y_0$ & $\FeH_0$ & $\alpha$ & $t/\mathrm{Gyr}$ & $a_3/10^{-6}\uHz$ & $a_{-1}/10^{-7}\uHz$ \\\midrule
Hare & &  & $1.723$ & $0.238$ & $0.147$ & $2.29$ & $2.42$ & $0$ \\
Hound & Fid.,~ & pre & $1.771 \pm 0.098$ & $0.207 \pm 0.040$ & $0.031 \pm 0.138$ & $2.14 \pm 0.15$ & $2.39 \pm 0.27$ & $-0.22 \pm 0.26$ \\
\\
\multirow{8}{*}{A}
 & Fid.,~ & pre & $1.671 \pm 0.067$ & $0.216 \pm 0.024$ & $0.069 \pm 0.065$ & $2.12 \pm 0.15$ & $2.88 \pm 0.27$ & $-1.67 \pm 0.40$ \\
 & Fid.,~ & post & $1.694 \pm 0.071$ & $0.213 \pm 0.022$ & $0.058 \pm 0.074$ & $2.09 \pm 0.16$ & $2.78 \pm 0.31$ & $-1.76 \pm 0.42$ \\
 & BG14-2,~ & pre & $1.732 \pm 0.083$ & $0.208 \pm 0.023$ & $0.075 \pm 0.064$ & $2.01 \pm 0.15$ & $2.71 \pm 0.28$ & $-2.46 \pm 0.51$ & $-3.40 \pm 0.56$ \\
 & BG14-2,~ & post & $1.761 \pm 0.083$ & $0.202 \pm 0.026$ & $0.059 \pm 0.085$ & $1.95 \pm 0.17$ & $2.64 \pm 0.29$ & $-2.73 \pm 0.62$ & $-3.48 \pm 0.79$ \\
 & $\alpha_\odot$,~ & pre & $1.700 \pm 0.080$ & $0.187 \pm 0.027$ & $0.075 \pm 0.074$ & $1.66 \pm 0.00$ & $3.35 \pm 0.30$ & $-3.05 \pm 0.27$ \\
 & $\alpha_\odot$,~ & post & $1.731 \pm 0.077$ & $0.185 \pm 0.024$ & $0.025 \pm 0.084$ & $1.66 \pm 0.00$ & $3.01 \pm 0.32$ & $-3.20 \pm 0.24$ \\
 & MR,~ & pre & $1.495 \pm 0.017$ & $0.269 \pm 0.012$ & $0.204 \pm 0.062$ & $2.17 \pm 0.17$ & $3.41 \pm 0.31$ & $-1.82 \pm 0.35$ \\
 & MR,~ & post & $1.497 \pm 0.018$ & $0.271 \pm 0.012$ & $0.213 \pm 0.053$ & $2.18 \pm 0.21$ & $3.39 \pm 0.23$ & $-1.87 \pm 0.45$ \\
\\
\multirow{8}{*}{B}
 & Fid.,~ & pre & $1.549 \pm 0.073$ & $0.200 \pm 0.023$ & $-0.361 \pm 0.072$ & $1.74 \pm 0.11$ & $2.78 \pm 0.28$ & $-6.23 \pm 0.86$ \\
 & Fid.,~ & post & $1.593 \pm 0.076$ & $0.175 \pm 0.027$ & $-0.257 \pm 0.100$ & $1.90 \pm 0.18$ & $3.26 \pm 0.50$ & $-4.30 \pm 1.49$ \\
 & BG14-2,~ & pre & $1.559 \pm 0.075$ & $0.202 \pm 0.026$ & $-0.322 \pm 0.075$ & $1.71 \pm 0.13$ & $2.78 \pm 0.31$ & $-8.04 \pm 1.42$ & $-7.45 \pm 2.72$ \\
 & BG14-2,~ & post & $1.596 \pm 0.080$ & $0.185 \pm 0.030$ & $-0.282 \pm 0.112$ & $1.80 \pm 0.20$ & $2.98 \pm 0.48$ & $-6.68 \pm 2.28$ & $-5.93 \pm 2.24$ \\
 & $\alpha_\odot$,~ & pre & $1.562 \pm 0.077$ & $0.193 \pm 0.022$ & $-0.348 \pm 0.067$ & $1.66 \pm 0.00$ & $2.86 \pm 0.27$ & $-6.72 \pm 0.62$ \\
 & $\alpha_\odot$,~ & post & $1.615 \pm 0.080$ & $0.169 \pm 0.031$ & $-0.319 \pm 0.101$ & $1.66 \pm 0.00$ & $3.06 \pm 0.43$ & $-6.25 \pm 0.68$ \\
 & MR,~ & pre & $1.439 \pm 0.034$ & $0.234 \pm 0.015$ & $-0.294 \pm 0.053$ & $1.80 \pm 0.09$ & $3.03 \pm 0.19$ & $-6.30 \pm 0.85$ \\
 & MR,~ & post & $1.443 \pm 0.029$ & $0.225 \pm 0.017$ & $-0.163 \pm 0.094$ & $1.95 \pm 0.16$ & $3.67 \pm 0.66$ & $-4.62 \pm 1.39$ \\
\\
\multirow{8}{*}{C}
 & Fid.,~ & pre & $1.270 \pm 0.064$ & $0.222 \pm 0.027$ & $-0.486 \pm 0.100$ & $1.68 \pm 0.09$ & $4.38 \pm 0.41$ & $-7.75 \pm 1.05$ \\
 & Fid.,~ & post & $1.379 \pm 0.064$ & $0.143 \pm 0.040$ & $-0.253 \pm 0.125$ & $1.89 \pm 0.18$ & $6.79 \pm 1.43$ & $-3.61 \pm 1.94$ \\
 & BG14-2,~ & pre & $1.313 \pm 0.061$ & $0.211 \pm 0.026$ & $-0.432 \pm 0.107$ & $1.57 \pm 0.13$ & $4.38 \pm 0.44$ & $-10.77 \pm 2.05$ & $-9.17 \pm 2.97$ \\
 & BG14-2,~ & post & $1.369 \pm 0.072$ & $0.164 \pm 0.041$ & $-0.301 \pm 0.117$ & $1.78 \pm 0.18$ & $5.76 \pm 1.09$ & $-6.17 \pm 2.70$ & $-4.58 \pm 2.27$ \\
 & $\alpha_\odot$,~ & pre & $1.277 \pm 0.048$ & $0.215 \pm 0.017$ & $-0.499 \pm 0.065$ & $1.66 \pm 0.00$ & $4.46 \pm 0.34$ & $-7.93 \pm 0.76$ \\
 & $\alpha_\odot$,~ & post & $1.384 \pm 0.062$ & $0.142 \pm 0.036$ & $-0.314 \pm 0.092$ & $1.66 \pm 0.00$ & $6.34 \pm 1.13$ & $-5.61 \pm 0.86$ \\
 & MR,~ & pre & $1.179 \pm 0.011$ & $0.257 \pm 0.012$ & $-0.369 \pm 0.071$ & $1.63 \pm 0.11$ & $4.95 \pm 0.27$ & $-8.41 \pm 1.10$ \\
 & MR,~ & post & $1.180 \pm 0.014$ & $0.243 \pm 0.019$ & $-0.175 \pm 0.118$ & $1.86 \pm 0.18$ & $6.58 \pm 1.21$ & $-5.37 \pm 1.76$ \\
\bottomrule
\end{tabular}

  \label{t:mdl}
\end{table*}

\begin{table*}
  \centering
  \caption{The derived stellar parameters for the model fits, as
    labelled in Table~\ref{t:mdl}.  The columns are the
    radius, mean density, surface gravity, surface metallicity,
    effective temperature, luminosity and $\chi^2$ per degree of
    freedom.}
  \begin{tabular}{cr@{}lr@{\,$\pm$\,}lr@{\,$\pm$\,}lr@{\,$\pm$\,}lr@{\,$\pm$\,}lr@{\,$\pm$\,}lr@{\,$\pm$\,}lc}
\toprule
Star & \twocell{Run} & \twocell{$R/R_\odot$} & \twocell{$\bar\rho/(10^{-3}\bar\rho_\odot)$} & \twocell{$\log g$} & \twocell{$\FeH_s$} & \twocell{$\Teff/\K$} & \twocell{$\log L/L_\odot$} & $\chi^2\st{red}$ \\\midrule
Hare & & & \twocell{$11.113$} & \twocell{$1.255$} & \twocell{$2.578$} & \twocell{$0.160$} & \twocell{$4792$} & \twocell{$1.771$} \\
Hound & Fid.,~ & pre & $11.218$ & $0.220$ & $1.255$ & $0.004$ & $2.582$ & $0.007$ & $0.045$ & $0.137$ & $4751$ & $88$ & $1.765$ & $0.038$ & 1.87 \\
\\
\multirow{8}{*}{A}
 & Fid.,~ & pre & $11.002$ & $0.155$ & $1.254$ & $0.003$ & $2.574$ & $0.005$ & $0.084$ & $0.065$ & $4708$ & $92$ & $1.732$ & $0.035$ & 3.31 \\
 & Fid.,~ & post & $11.064$ & $0.165$ & $1.251$ & $0.003$ & $2.575$ & $0.005$ & $0.072$ & $0.074$ & $4704$ & $89$ & $1.736$ & $0.034$ & 3.31 \\
 & BG14-2,~ & pre & $11.099$ & $0.184$ & $1.267$ & $0.003$ & $2.582$ & $0.006$ & $0.089$ & $0.063$ & $4648$ & $88$ & $1.717$ & $0.035$ & 2.45 \\
 & BG14-2,~ & post & $11.170$ & $0.183$ & $1.263$ & $0.004$ & $2.583$ & $0.006$ & $0.074$ & $0.085$ & $4622$ & $98$ & $1.713$ & $0.037$ & 2.39 \\
 & $\alpha_\odot$,~ & pre & $11.066$ & $0.184$ & $1.254$ & $0.004$ & $2.576$ & $0.006$ & $0.090$ & $0.074$ & $4409$ & $23$ & $1.623$ & $0.022$ & 3.94 \\
 & $\alpha_\odot$,~ & post & $11.147$ & $0.177$ & $1.250$ & $0.004$ & $2.578$ & $0.006$ & $0.041$ & $0.083$ & $4445$ & $33$ & $1.644$ & $0.025$ & 3.75 \\
 & MR,~ & pre & $10.580$ & $0.043$ & $1.262$ & $0.001$ & $2.559$ & $0.002$ & $0.218$ & $0.062$ & $4701$ & $105$ & $1.696$ & $0.039$ & 3.92 \\
 & MR,~ & post & $10.590$ & $0.046$ & $1.260$ & $0.001$ & $2.559$ & $0.002$ & $0.227$ & $0.052$ & $4705$ & $122$ & $1.698$ & $0.045$ & 4.10 \\
\\
\multirow{8}{*}{B}
 & Fid.,~ & pre & $13.661$ & $0.229$ & $0.608$ & $0.002$ & $2.351$ & $0.006$ & $-0.344$ & $0.072$ & $4596$ & $84$ & $1.880$ & $0.037$ & 2.93 \\
 & Fid.,~ & post & $13.807$ & $0.231$ & $0.605$ & $0.002$ & $2.354$ & $0.006$ & $-0.240$ & $0.099$ & $4610$ & $85$ & $1.894$ & $0.034$ & 2.93 \\
 & BG14-2,~ & pre & $13.617$ & $0.228$ & $0.618$ & $0.004$ & $2.357$ & $0.007$ & $-0.306$ & $0.074$ & $4564$ & $92$ & $1.864$ & $0.036$ & 2.49 \\
 & BG14-2,~ & post & $13.750$ & $0.242$ & $0.614$ & $0.003$ & $2.359$ & $0.007$ & $-0.266$ & $0.111$ & $4578$ & $98$ & $1.879$ & $0.040$ & 2.70 \\
 & $\alpha_\odot$,~ & pre & $13.699$ & $0.239$ & $0.607$ & $0.002$ & $2.352$ & $0.006$ & $-0.331$ & $0.066$ & $4530$ & $22$ & $1.857$ & $0.021$ & 2.75 \\
 & $\alpha_\odot$,~ & post & $13.872$ & $0.242$ & $0.605$ & $0.002$ & $2.356$ & $0.006$ & $-0.302$ & $0.100$ & $4499$ & $51$ & $1.856$ & $0.028$ & 2.95 \\
 & MR,~ & pre & $13.311$ & $0.113$ & $0.610$ & $0.001$ & $2.342$ & $0.003$ & $-0.278$ & $0.053$ & $4616$ & $64$ & $1.865$ & $0.026$ & 3.47 \\
 & MR,~ & post & $13.332$ & $0.094$ & $0.609$ & $0.001$ & $2.342$ & $0.003$ & $-0.148$ & $0.093$ & $4615$ & $80$ & $1.866$ & $0.032$ & 3.74 \\
\\
\multirow{8}{*}{C}
 & Fid.,~ & pre & $13.392$ & $0.240$ & $0.529$ & $0.002$ & $2.281$ & $0.007$ & $-0.469$ & $0.099$ & $4571$ & $74$ & $1.854$ & $0.037$ & 4.26 \\
 & Fid.,~ & post & $13.793$ & $0.227$ & $0.525$ & $0.002$ & $2.291$ & $0.006$ & $-0.234$ & $0.125$ & $4506$ & $82$ & $1.855$ & $0.035$ & 3.66 \\
 & BG14-2,~ & pre & $13.466$ & $0.222$ & $0.537$ & $0.003$ & $2.291$ & $0.007$ & $-0.415$ & $0.106$ & $4469$ & $102$ & $1.819$ & $0.042$ & 3.65 \\
 & BG14-2,~ & post & $13.712$ & $0.258$ & $0.531$ & $0.003$ & $2.293$ & $0.007$ & $-0.283$ & $0.117$ & $4486$ & $83$ & $1.842$ & $0.037$ & 3.61 \\
 & $\alpha_\odot$,~ & pre & $13.418$ & $0.179$ & $0.529$ & $0.001$ & $2.282$ & $0.005$ & $-0.481$ & $0.065$ & $4560$ & $23$ & $1.851$ & $0.018$ & 4.01 \\
 & $\alpha_\odot$,~ & post & $13.812$ & $0.218$ & $0.525$ & $0.001$ & $2.292$ & $0.006$ & $-0.294$ & $0.092$ & $4404$ & $60$ & $1.816$ & $0.026$ & 3.64 \\
 & MR,~ & pre & $13.045$ & $0.041$ & $0.531$ & $0.001$ & $2.272$ & $0.001$ & $-0.353$ & $0.071$ & $4504$ & $86$ & $1.806$ & $0.033$ & 4.52 \\
 & MR,~ & post & $13.059$ & $0.054$ & $0.530$ & $0.001$ & $2.271$ & $0.002$ & $-0.158$ & $0.118$ & $4521$ & $89$ & $1.813$ & $0.034$ & 4.57 \\
\bottomrule
\end{tabular}

  \label{t:inf}
\end{table*}

\begin{figure*}
  \centering
\begin{tikzpicture}
  \node[inner sep=0pt] (img) at (0,0)
       {\includegraphics[width=2\columnwidth]{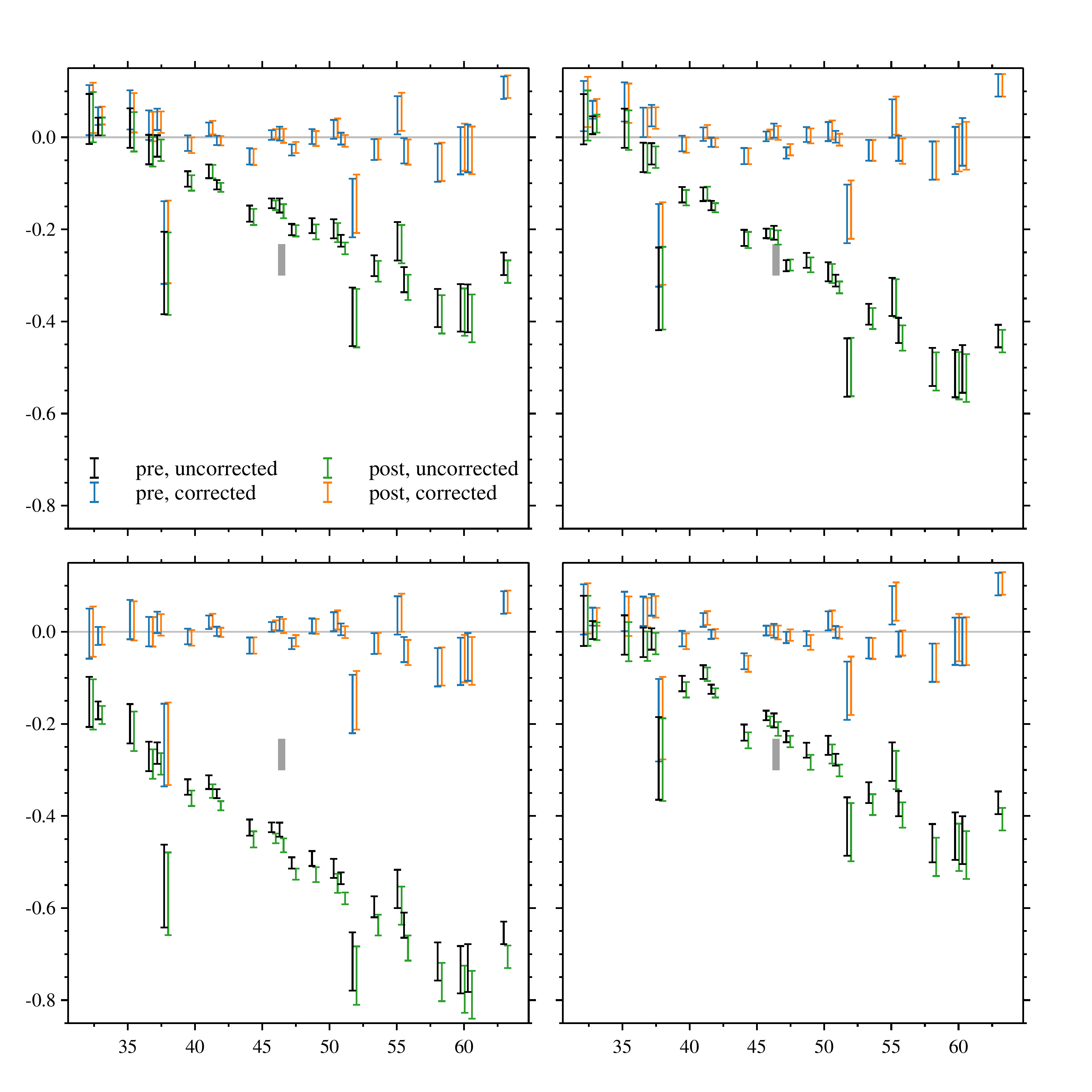}};
  \node (x) at ($(img.south)!0.02!(img.north)$) {\Large $\nu\st{obs}/\uHz$};
  \node[rotate=90] (y) at ($(img.west)!0.0!(img.east)$) {\Large $(\nu\st{obs}-\nu)/\uHz$};
  \node (z1) at (-6.2, 2.2) {\Large Fiducial};
  \node (z2) at ( 1.0, 2.2) {\Large $\alpha_\odot$};
  \node (z3) at (-6.2,-6.2) {\Large BG14-2};
  \node (z4) at ( 1.0,-6.2) {\Large MR};
 \end{tikzpicture}
 \caption{Frequency differences before and after applying a surface
   correction as a function of observed frequency for best-fitting
   models of star A (\starA{}).  The four panels are for the fiducial
   fit (top left), the fit with solar-calibrated mixing-length (top
   right), the fit using the two-term surface correction by
   \citet{ball2014} (bottom left) and the fit using the orbital mass
   and radius as observable constraints (bottom right).  For each fit,
   we have plotted the frequency difference before (uncorrected) and
   after correction (corrected) for both the pre- and post-RGB bump
   models.  The post-RGB bump frequencies are shifted right
   by $0.3\uHz$ for clarity.
   The solid grey bar indicates the surface correction
     predicted by eq.~(10) of \citet{sonoi2015}.}
 \label{f:best_841}
\end{figure*}
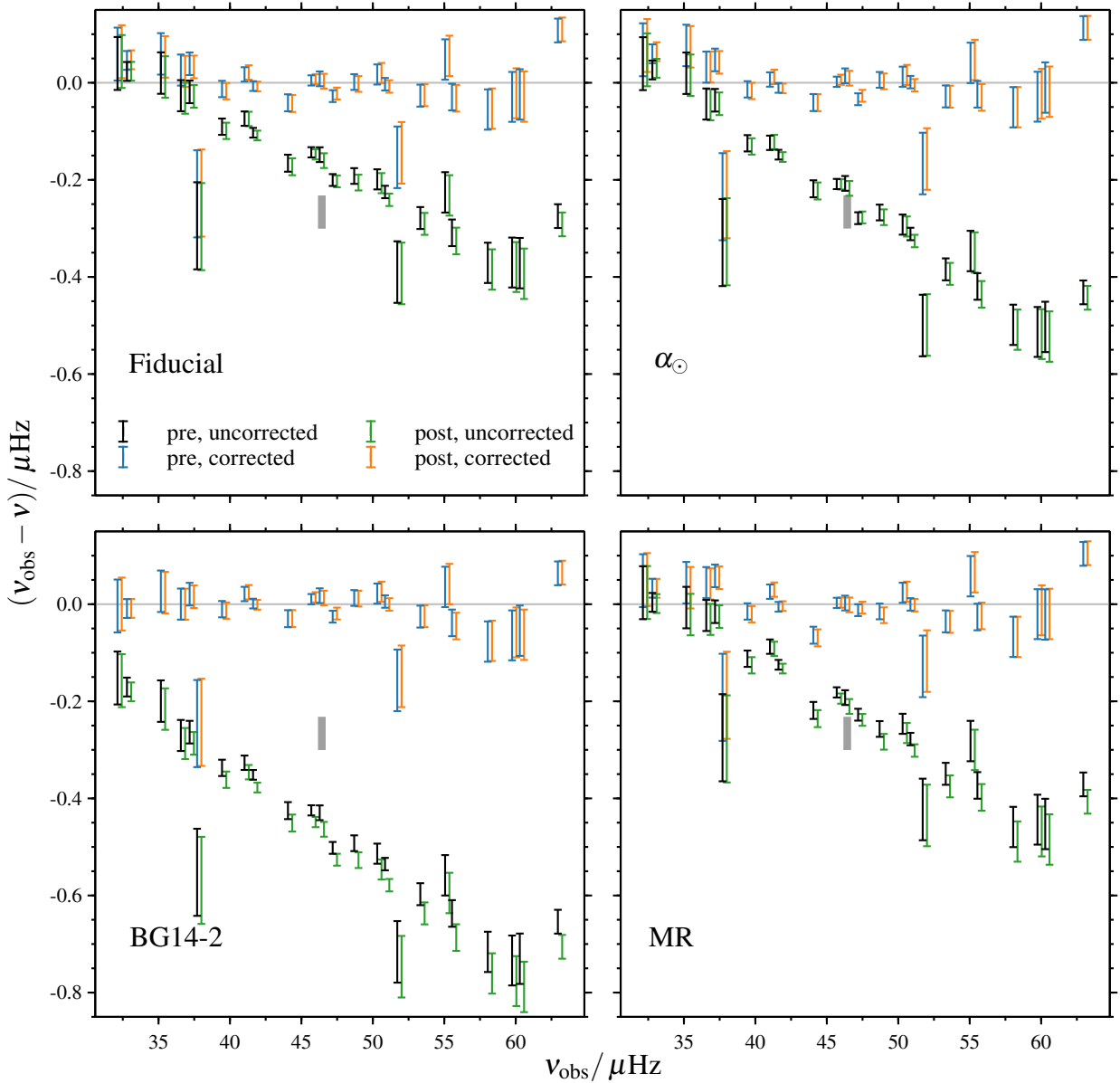

\begin{figure*}
  \centering
\begin{tikzpicture}
  \node[inner sep=0pt] (img) at (0,0)
       {\includegraphics[width=2\columnwidth]{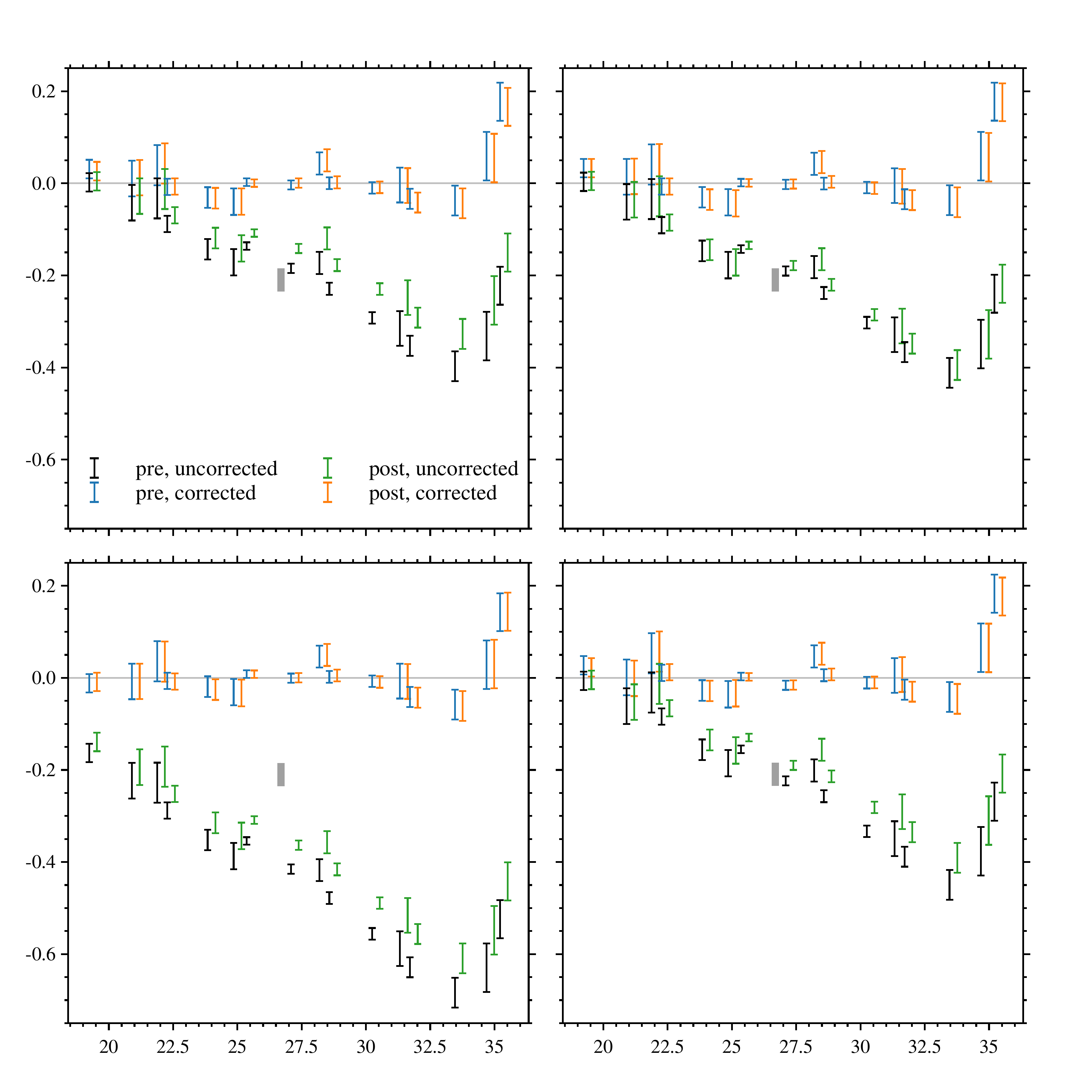}};
  \node (x) at ($(img.south)!0.02!(img.north)$) {\Large $\nu\st{obs}/\uHz$};
  \node[rotate=90] (y) at ($(img.west)!0.0!(img.east)$) {\Large $(\nu\st{obs}-\nu)/\uHz$};
  \node (z1) at (-6.2, 2.2) {\Large Fiducial};
  \node (z2) at ( 1.0, 2.2) {\Large $\alpha_\odot$};
  \node (z3) at (-6.2,-6.2) {\Large BG14-2};
  \node (z4) at ( 1.0,-6.2) {\Large MR};
 \end{tikzpicture}
 \caption{As in Fig.~\ref{f:best_841} for star B (\starB{}).}
 \label{f:best_954}
\end{figure*}

\begin{figure*}
  \centering
\begin{tikzpicture}
  \node[inner sep=0pt] (img) at (0,0)
       {\includegraphics[width=2\columnwidth]{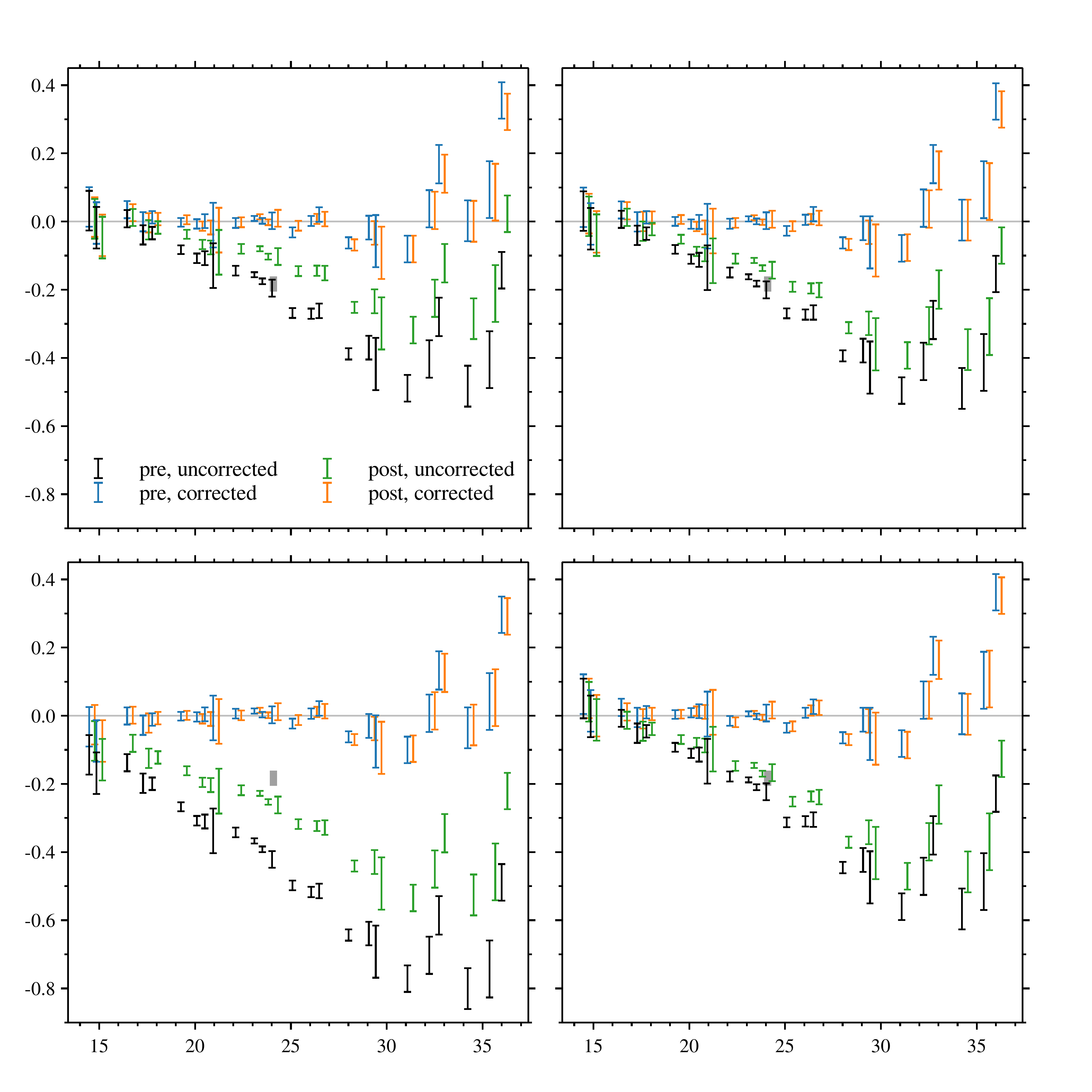}};
  \node (x) at ($(img.south)!0.02!(img.north)$) {\Large $\nu\st{obs}/\uHz$};
  \node[rotate=90] (y) at ($(img.west)!0.0!(img.east)$) {\Large $(\nu\st{obs}-\nu)/\uHz$};
  \node (z1) at (-6.2, 2.2) {\Large Fiducial};
  \node (z2) at ( 1.0, 2.2) {\Large $\alpha_\odot$};
  \node (z3) at (-6.2,-6.2) {\Large BG14-2};
  \node (z4) at ( 1.0,-6.2) {\Large MR};
 \end{tikzpicture}
 \caption{As in Fig.~\ref{f:best_841} for star C (\starC{}).}
 \label{f:best_564}
\end{figure*}

\section{Results}
\label{s:results}

\subsection{Overview}

The best-fitting model parameters are presented in Tables~\ref{t:mdl}
and \ref{t:inf}.  Table~\ref{t:mdl} provides the model parameters that
specify the stellar model while Table~\ref{t:inf} provides other
observable and derived quantities.  Each row specifies which of the
four fits is given and the final ``pre'' or ``post'' indicates whether
that fit is for models before or after the RGB bump.
Figs~\ref{f:best_841}--\ref{f:best_564} show the differences between
the observed and modelled frequencies for each star before and after
the surface effects have been corrected, with error bars
indicating the observed uncertainties.  In each figure,
the four fits appear
in different panels with the same $x$- and $y$-axes.
We have also indicated the surface correction predicted by eq.~(10) of
\citet{sonoi2015}.

The first point the reader may notice in Tables~\ref{t:mdl} and
\ref{t:inf} is that the uncertainties are larger than often quoted in
the asteroseismology of dwarf solar-like oscillators---especially
Sun-like stars---when individual frequencies are available
\citep[e.g.][]{reese2016,legacy2,bellinger2017kasc}.
Our uncertainties are also larger than those given for the
  slightly less evolved red giants ($\log g\approx 3$) studied
  by \citet{perez2016}.
Though our
uncertainties on the radii are around the 2 per cent level, our
uncertainties on the masses can be nearly 6 per cent.  This is at least
in part because our suppression of g-modes in the core discards a
great deal of information about the stellar interior.  For example, we
cannot possibly have the diagnostic power of the period spacing that
has been demonstrated by \citet{mosser2014}.  We are, however, still
able to tightly constrain the mean density and surface gravity, and we
achieve age uncertainties of about 10 to 20 per cent.  Above all, we
are still able to make useful inferences about the surface effects,
which are our main interests here.

We also note that the fits that are not constrained by the orbital
solution (i.e. run MR) have initial helium abundances $Y_0$ that are
smaller---sometimes significantly---than the primordial value of
$0.249^{+0.025}_{-0.026}$ \citep{planck_bbn2016}.

\subsection{Two solutions}

As mentioned above, for most choices of initial model parameters we
found similarly good models both before and after the RGB bump.  In
both cases, the star is ascending the giant branch. i.e. the
luminosity is increasing and the effective temperature decreasing.
For some tracks the code recorded models during the RGB bump (when the
luminosity decreases and the effective temperature increases) but such
models always provided much poorer fits, unlike the
pre- and post-bump models we report.  In stars A and B the two
models are of similar quality, though in star C the post-bump
model appears to fit the data slightly better.

The occurrence of two solutions is presumably also a
consequence of suppressing the g-modes in the core.  For a given
choice of initial parameters, both models will have similar p-mode
qualities like surface gravity and mean density, and hence only a
modest difference in their p-mode spectra.  The full models should
differ in their g-mode spectra but we have in effect discarded this
information.

The existence of the two solutions does not undermine the quantitative
results.  In stars A and B, all of the pre- and post-bump fits are
consistent within uncertainties, even for the mean densities
$\bar\rho$ and surface gravities $\log g$, which have small
uncertainties.  This is not so for star C and, given that
the post-bump models fit better in most cases,
we conclude that star C has evolved past the RGB bump.

\begin{figure*}
  \centering
\begin{tikzpicture}
  \node[inner sep=0pt] (img) at (0,0)
       {\includegraphics[width=2\columnwidth]{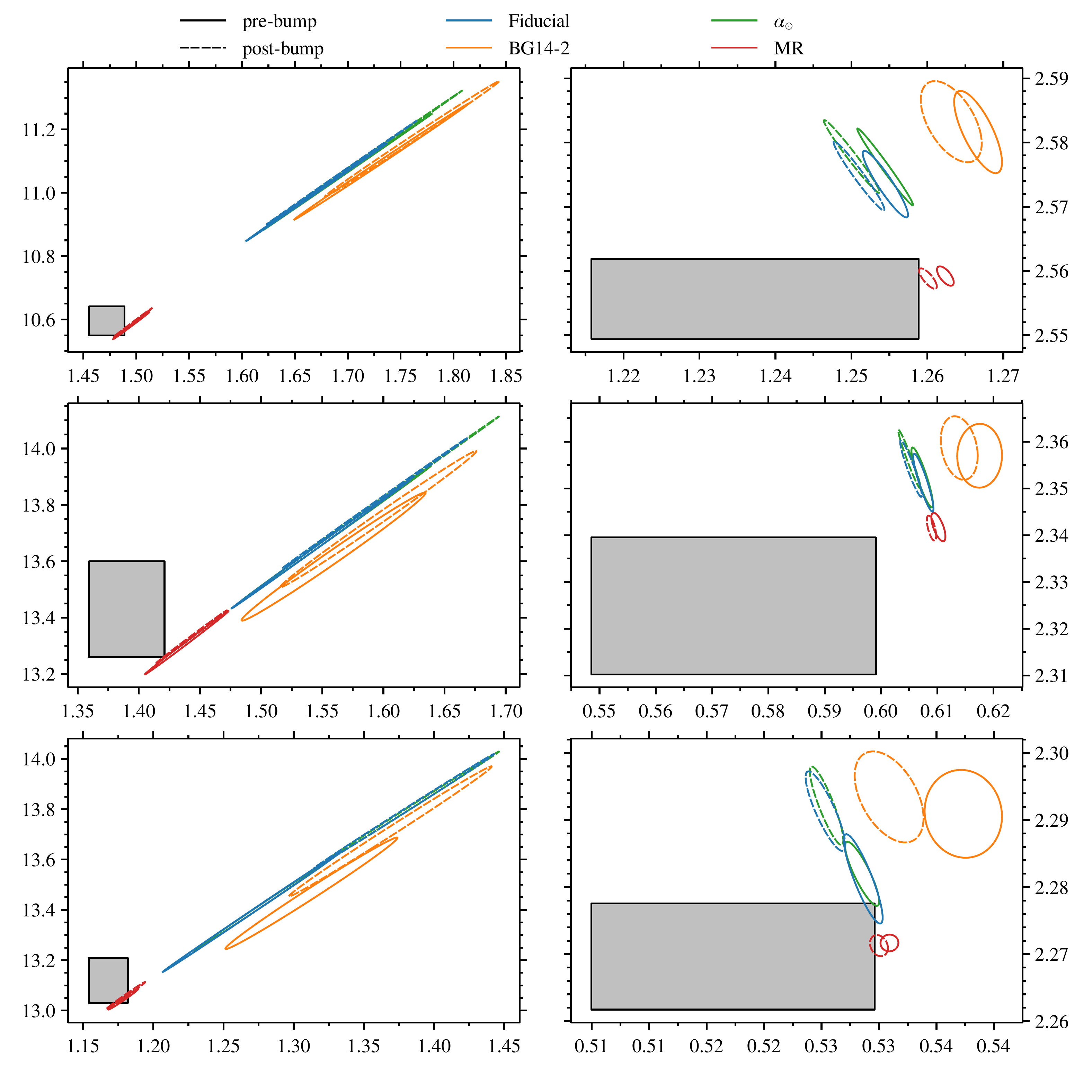}};
  \node (x) at ($(img.south)!0.5!(img.south east)!0.02!(img.north)$)
        {\Large $\bar\rho/(10^{-3}\bar\rho_\odot)$};
  \node (x) at ($(img.south)!0.5!(img.south west)!0.02!(img.north)$)
        {\Large $M/M_\odot$};
  \node (A) at (-6.5, 6.5) {\Huge A};
  \node (B) at (-6.5, 1.25) {\Huge B};
  \node (C) at (-6.5,-4.) {\Huge C};
  \node[rotate=90] (y) at ($(img.west)!0.0!(img.east)$) {\Large $R/R_\odot$};
  \node[rotate=-90] (y) at ($(img.west)!1.0!(img.east)$) {\Large $\log g$};
 \end{tikzpicture}
 \caption{Plot showing the masses $M$ versus radii $R$ (left) and
   mean densities $\bar\rho$ versus surface gravities $\log g$ (right) for the various fits,
   compared to the values derived from the orbital solutions (grey boxes).  From top to bottom,
   the plots show the results for stars A, B and C
   (\starA{}, \starB{} and \starC{}).  The blue, orange,
   green and red ellipses correspond to the fiducial fits, the fits
   with the two-term correction by \citet{ball2014}, the fits with a
   solar-calibrated mixing-length parameter, and the fits with the
   mass and radius constraints from the orbital solutions.  In each
   case, the parameters of the best fitting models before and after
   the RGB bump are shown with solid (pre-bump) or dashed (post-bump)
   lines.  Note that in some cases the pre- and post-bump solutions
   are indistinguishable.
 }
 \label{f:MRrhologg}
\end{figure*}

\subsection{Comparison with binary solutions}

In Fig.~\ref{f:MRrhologg}, we compare the masses, radii, mean
densities and surface gravities of the three stars with the
results obtained from the orbital solutions of the three stars.
Without constraining the masses and radii to match the values found in
the orbital solutions, the models generally disagree with the orbital
parameters at roughly the $2$ to $3\sigma$ level.

For all three stars, the best-fitting models obtained using the
  oscillations are larger and more massive than the radii and masses
  derived from the binary solutions.
These discrepancies
are similar to those found by \citet{gaulme2016} when using scaling
relation.
\citet{themessl2018} finds the same discrepancies when not
accounting for the mass, temperature, metallicity or surface effects.
The literature contains extensive
discussion about how accurate and precise the scaling relations are
and some propose corrections to the reference values 
  based on stellar models \citep[e.g.][]{guggenberger2016, guggenberger2017}.
  The results here suggest that such corrections might be
  influenced by inaccuracies in the stellar models themselves, which can be
  studied in systems that have independent constraints on mass or
radius, like the binaries studied here.

It is interesting that the
model parameter that changes most when constraining the fits by the
dynamical masses and radii is the initial helium abundance $Y_0$.
The correlation between the mass $M$ and $Y_0$ is
  well-known \citep[e.g.][]{lebreton2014}.
  The initial helium abundance $Y_0$
directly affects the mean density of the model and
it appears that the increased $Y_0$ means that the mean densities of
the stellar models are much less affected by the inclusion of the
orbital constraints.  Increasing $Y_0$ would also make our models more
similar to those of \citet{li2018}, who computed models with a fixed
enrichment law with $Y_0=0.249+1.33Z_0$, where $Z_0$ is the model's
initial metal content.

We note again that these results are based on models that
  neglect gravitational settling and radiative levitation.
Gravitational settling could play a role by changing the mean
molecular weight and therefore the density throughout the star.
However, without an opposing
process like rotation or radiative levitation,
current implementations of gravitational settling make unrealistic
predictions about stellar properties on the main sequence.  The extent
of convective core overshooting is another poorly-constrained process in stellar
modelling that would change the stars' mean
densities.  Thus, there remain several processes that might influence
these results, which introduces the possibility that
further studies like this one
might be able to constrain those same processes.

\begin{table}
  \centering
  \caption{Comparison of surface effect magnitudes between the
    different fits and the predictions of \citet{sonoi2015}.  Values
    are $1000$ times the fractional surface correction at
    $\nu\st{max}$.}  \begin{tabular}{r@{}lr@{}lr@{}lr@{}l}
\toprule
& & \multicolumn{6}{c}{Star} \\
& & \twocell{A} & \twocell{B} & \twocell{C} \\
\midrule
\twocell{\citeauthor{sonoi2015}, eq.~(10)} & $-5.74$\, & $\pm$\,$0.74$ & $-7.86$\, & $\pm$\,$0.95$ & $-7.61$\, & $\pm$\,$0.94$\\
\twocell{\citeauthor{sonoi2015}, eq.~(21)} & $-5.68$\, & $\pm$\,$0.60$ & $-8.47$\, & $\pm$\,$0.84$ & $-8.91$\, & $\pm$\,$0.89$ \\
Fid.,~ &  pre & $-3.58$\, & $\pm$\,$0.72$ & $-8.18$\, & $\pm$\,$0.92$ & $-8.72$\, & $\pm$\,$0.98$ \\
Fid.,~ &  post & $-3.75$\, & $\pm$\,$0.69$ & $-6.44$\, & $\pm$\,$1.58$ & $-5.19$\, & $\pm$\,$2.19$ \\
$\alpha_\odot$,~ &  pre & $-5.50$\, & $\pm$\,$0.51$ & $-8.57$\, & $\pm$\,$0.81$ & $-8.88$\, & $\pm$\,$0.80$ \\
$\alpha_\odot$,~ &  post & $-5.71$\, & $\pm$\,$0.46$ & $-8.21$\, & $\pm$\,$0.79$ & $-7.12$\, & $\pm$\,$0.90$ \\
BG14-2,~ &  pre & $-9.94$\, & $\pm$\,$1.23$ & $-17.23$\, & $\pm$\,$3.50$ & $-18.99$\, & $\pm$\,$3.63$ \\
BG14-2,~ &  post & $-10.37$\, & $\pm$\,$1.63$ & $-14.91$\, & $\pm$\,$3.69$ & $-12.07$\, & $\pm$\,$4.08$ \\
MR,~ &  pre & $-4.07$\, & $\pm$\,$0.60$ & $-8.54$\, & $\pm$\,$0.93$ & $-9.44$\, & $\pm$\,$0.97$ \\
MR,~ &  post & $-4.21$\, & $\pm$\,$0.74$ & $-7.14$\, & $\pm$\,$1.49$ & $-7.30$\, & $\pm$\,$1.64$ \\
\bottomrule
\end{tabular}

  \label{t:sonoi}
\end{table}

\subsection{Surface effects}

The overall scale of the surface corrections is roughly consistent for
all the models that use the one-term (cubic) fit by \citet{ball2014}.
For the two-term (combined) fit, the surface correction is somewhat
larger in magnitude for all three stars.  For stars A and B, the scale
of the correction is roughly the same for the pre- and post-bump
models, as we expect because the surface effect is determined by near
surface properties that are similar in both cases.  The surface
effects differ between pre- and post-bump models for star C though
there the post-bump solution is preferred.

In Table~\ref{t:sonoi} we compare the surface correction at
$\nu\st{max}$ with the predictions by \citet{sonoi2015} computed using
the observed data, with uncertainties determined by making random
realisations of the observations.  We note that the comparison is far
from exact.  The results in \citet{sonoi2015} are only for the part of
the surface effect caused by the poor-modelling of the background
stellar model and do not incorporate any changes to the dynamics of
the oscillations.  It does, however, still give some idea of the kind
of surface effect that is to be expected.  As shown by the figures,
the one-term (cubic) correction gives results that are similar to
the predicted corrections, whereas the two-term (combined) fit is
usually larger.

One of the main results of this article, then, is that fitting models
to data using the one-term correction by \citet{ball2014} leads to
surface corrections that are similar to those predicted by
\citet{sonoi2015}.  Though this could only mean that both models are
equally wrong, it is at least encouraging that this result is achieved
without additional constraints.  The results are also consistent with
the surface corrections determined by \citet{perez2016}, who found
a relative frequency difference
$\delta\nu/\nu\approx-0.005\nu\st{max}$ at $\nu\st{max}$.

\begin{table}
  \centering
  \caption{Mixing-length parameters determined by interpolating in the
    results of \citet{magic2015}.  The second column gives the
    predicted mixing-length parameter relative to the solar value in
    the grid.  The third column is the same value multiplied by our
    solar-calibrated value of $\alpha_\odot=1.66$, for comparison with
    Table~\ref{t:mdl}.}  \begin{tabular}{ccc}
\toprule
Star &
$\alpha/\alpha_{\odot,\textsc{Stagger}}$ &
$\times\,\alpha_{\odot,\mathrm{MESA}}$ \\
\midrule
A & $0.959  \pm 0.009$ & $1.591  \pm 0.015$\\
B & $0.951  \pm 0.008$ & $1.579  \pm 0.014$\\
C & $0.951  \pm 0.008$ & $1.578  \pm 0.013$\\
\bottomrule
\end{tabular}
  \label{t:alpha}
\end{table}

\subsection{Mixing-length parameters}

Three of our fits for each star allowed the mixing-length parameter
$\alpha$ to vary freely.  Although there is no obvious trend between
the different stars, all the values are larger than the
solar-calibrated value $\alpha_\odot=1.66$.  \citet{ball2017} also
found super-solar mixing-length parameters for their best-fitting
models, in that case for fits to subgiants and low-luminosity red
giants ($3.5\lesssim\log g\lesssim3.8$), as did \citet{li2018} for
their sample of six eclipsing binaries with similar parameters to ours
(and including stars A and B).  Similarly, \citet{tayar2017} found
that, to reconcile temperatures from spectroscopy and evolutionary
tracks, they too would need to increase the mixing-length parameters
of their models above the solar-calibrated values.

Like \citet{ball2017}, we have compared our best-fitting mixing-length
parameters with the predictions of \citet{magic2015}, which are based
on calibrations of mixing-length envelope models to three-dimensional
radiation hydrodynamics simulations from the \textsc{Stagger} code.
We generated $10^5$ values of $\nu\st{max}$, $\Teff$ and $\FeH$ and
used them to interpolate in their data for mixing-length parameters
calibrated to the entropy jump at the bottom boundary of the
simulation.  In Table~\ref{t:alpha}, we list the means and standard
deviations determined for each star relative to the solar-calibrated
value in the simulation data.  We also give the value multiplied by
the solar-calibrated mixing-length parameter for the MESA models
presented here.

Table~\ref{t:alpha} shows that the simulations support smaller
mixing-length parameters than our best-fitting stellar models.  In
fact, they suggest that the mixing-length parameter should be less
than the solar-calibrated value.  Combined with the results of
\citet{ball2017}, \citet{li2018} and \citet{tayar2017}, there appears
to be a growing tension with the predictions of hydrodynamics
simulations, or at least those of \citet{magic2015}.  It is not clear
what the source of such tension might be.  One possibility is the
choice of atmospheric model, which is known to affect the
mixing-length parameter.  In that case, the implication would be that
the Eddington approximation gives a poor atmospheric model for evolved
stars.  This could be tested by incorporating the atmospheric models
of the hydrodynamics simulations, which could be performed using the
method set out by \citet{trampedach2014a}, though this is
beyond the scope of the present work.

\section{Conclusion}
\label{s:conclusion}

The mode suppression method we have presented gives unbiased results
at the cost of greater uncertainties in the inferred parameters.  With
this level of uncertainty, however, we still find that the one-term
(cubic) surface correction by \citet{ball2014} leads to a correction
that agrees with the prediction by \citet{sonoi2015}.  This is broadly
true for the solutions before or after the RGB bump and whether or not
the dynamical masses and radii are included as constraints.  The
two-term (combined) correction by \citet{ball2014} appears to lead to
somewhat larger surface corrections.  Overall, the surface
  effects in the three stars are robust, with shifts of about
  $0.1$--$0.3\uHz$ at $\nu\st{max}$ for all three stars across all the fits
  that use the cubic correction.

Without constraining them to agree, the stellar models lead to
significantly discrepant masses and radii compared with the orbital
solutions.  The masses and radii are still incorrect even when the
mixing-length parameter is fixed at the solar-calibrated value.
The discrepancy in all the stars is
similar to the roughly 15 and 5 per cent in mass and radius found by
\citet{gaulme2016} and \citet{themessl2018} when the scaling relations
are not corrected for the effects of mass, effective temperature, metallicity
or surface effects.  The initial helium abundance $Y_0$ changes the
most when the masses and radii are constrained to match the dynamical
values, which suggests that the difference might be caused by the
composition profile of the stellar models through processes like
gravitational settling and rotation, which have been ignored here.

While our mode suppression method has allowed us to infer stellar
properties and surface corrections for these three stars, it would
clearly be better to be able to exploit the full set of observed
modes.  As it is, we do not make use of any information about the
stellar core that is contained in the mixed modes and constrain
processes like core overshooting on the main sequence
\citep[e.g.][]{montalban2013, lagarde2016}.  Nevertheless, it
presents a step forward in establishing that the surface effect in RGB
stars manifests itself in a way similar to main-sequence stars and in
line with our expectations from hydrodynamic simulations.

\section*{Acknowledgements}

WHB acknowledges the support of the UK Science and Technology
Facilities Council (STFC).
The research leading to the presented results has received funding
from the European Research Council under the European Community's
Seventh Framework Programme (FP7/2007-2013) / ERC grant agreement no
338251 (StellarAges).

\bibliographystyle{../aa}
\bibliography{../master}

\appendix

\section{Source code for suppressing core gravity modes}
\label{a:nog}

\new{Here, we detail the modifications to MESA's Fortran source
  (revision 9575)
  necessary to reproduce the g-mode suppression described in
  Sec.~\ref{ss:nog}.  All the modifications are applied when the
  stellar model data is stored for ADIPLS, in the subroutine
  \texttt{store\_model\_for\_adipls} on lines 231--340 of the file
  \texttt{\$MESA\_DIR/star/astero/src/adipls\_support.f}.  Here,
  \texttt{\$MESA\_DIR} is the environment variable specifying MESA's
  top-level directory.}

\new{First, we declare a new set of variables at the beginning of the
  subroutine by inserting the following code at line 247.}

\begin{lstlisting}
  integer :: k
  real(dp) :: dlnP_dlnr, dlnrho_dlnr
\end{lstlisting}
\new{Then, we insert the following code at line 312 (in the unmodified
  file) to loop over the stellar model and perform the calculations
  described in Sec.~\ref{ss:nog}.}
\begin{lstlisting}
  ! suppress oscillations in the core by modifying AA directly
  if (dbg) write(*,*) 'modifying aa'
  do k=2,nn-1
    if ((x(k) < 0.4) .and. (aa(4,k) > 0)) then
      dlnP_dlnr = -aa(3,k)*aa(2,k)
      dlnrho_dlnr = -aa(2,k)-aa(4,k)
      aa(4,k) = 0d0                   ! A = 0
      aa(2,k) = aa(2,k)*aa(3,k)       ! multiply Vg by Gamma_1
      aa(3,k) = dlnP_dlnr/dlnrho_dlnr ! Gamma_1 = dlnP/dlnrho
      aa(2,k) = aa(2,k)/aa(3,k)       ! divide to get new Vg
    end if
  end do
  aa(4,1) = 2*aa(4,2)-aa(4,3)
  if (dbg) write(*,*) 'done modifying aa'
\end{lstlisting}
\new{Line 4 selects those points in the core that are convectively
  stable.  The hardcoded fractional radius $x<0.4$ can be any value in
  the convective envelope.  Lines 5 and 6 recover $\mathrm{d}\ln
  P/\mathrm{d}\ln r$ and $\mathrm{d}\ln \rho/\mathrm{d}\ln r$ from the
  existing model data.  Line 7 assigns $N^2=0$ and lines 8--10
  reassign the adiabatic index $\Gamma_1$ to its new value given by
  eq.~(\ref{e:newG1}).  Line 13 assigns $N^2$ at the central point by
  linear interpolation.}

\label{lastpage}
\end{document}